\documentstyle{BSAXwork}

\input epsf.sty

\def \AAP #1 #2 {{\em Astron. Astrophys.\/} {\bf #1}, #2}
\def \AAL #1 #2 {{\em Astron. Astrophys. Lett.\/} {\bf #1}, L#2}
\def \AAR #1 #2 {{\em Astron. Astrophys. Rev.\/} {\bf #1}, #2}
\def \AAS #1 #2 {{\em Astron. Astrophys. Suppl. Ser.\/} {\bf #1}, #2}
\def \AJ #1 #2 {{\em Astron. J.\/} {\bf #1}, #2}
\def \ANNREV #1 #2 {{\em Ann. Rev. Astron. Astrophys.\/} {\bf #1}, #2}
\def \APJ #1 #2 {{\em Astrophys. J.\/} {\bf #1}, #2}
\def \APJL #1 #2 {{\em Astrophys. J. Lett.\/} {\bf #1}, L#2}
\def \APJS #1 #2 {{\em Astrophys. J. Suppl.\/} {\bf #1}, #2}
\def \APSS #1 #2 {{\em Astrophys. Space Sci.\/} {\bf #1}, #2}
\def \ASR #1 #2 {{\em Adv. Space Res.\/} {\bf #1}, #2}
\def \BAIC #1 #2 {{\em Bull. Astron. Inst. Czechosl.\/} {\bf #1}, #2}
\def \JSQRT #1 #2 {{\em J. Quant. Spectrosc. Radiat. Transfer\/} {\bf #1}, #2}
\def \MN #1 #2 {{\em Mon. Not. R. Astr. Soc.\/} {\bf #1}, #2}
\def \MEM #1 #2 {{\em Mem. R. Astr. Soc.\/} {\bf #1}, #2}
\def \PLR #1 #2 {{\em Phys. Lett. Rev.\/} {\bf #1}, #2}
\def \PASJ #1 #2 {{\em Publ. Astron. Soc. Japan\/} {\bf #1}, #2}
\def \PASP #1 #2 {{\em Publ. Astr. Soc. Pacific\/} {\bf #1}, #2}
\def \NAT #1 #2 {{\em Nature\/} {\bf #1}, #2}
\def \SAIT #1 #2 {{\em Mem.\ Soc.\ Astron.\ It.\/} {\bf #1}, #2}
\def \MESS #1 #2 {{\em The Messenger\/} {\bf #1}, #2}
\def \ASTRNACH #1 #2 {{\em Astron. Nach.\/} {\bf #1}, #2}
\def \sax  {{\it Beppo}SAX~}
\def \gsim { \lower .75ex \hbox{$\sim$} \llap{\raise .27ex \hbox{$>$}} }
\def \lsim { \lower .75ex\hbox{$\sim$} \llap{\raise .27ex \hbox{$<$}} }

\begin{opening}

\title{The BeppoSAX view of extreme BL Lacs}
\author{L. Costamante$^1$, G. Ghisellini$^1$, A. Celotti$^2$, A. Wolter$^1$ }
\institute{$^1$Osservatorio Astronomico di Brera, Milano, Italy\\
$^2$SISSA/ISAS, Trieste, Italy}
\date{} 
\end{opening}

\begin{document}

\oddpagefooter{}{}{} 
\evenpagefooter{}{}{} 
\medskip  

\begin{abstract} 
We summarize the results of an observational program performed with the 
satellite \sax with the aim to find and study more extreme 
BL Lac objects. We  discuss the SEDs of the observed objects and
their impact on the ``blazar sequence" scenario, and consider
their relevance as possible TeV emitting sources.

\end{abstract}

\medskip

\section{Introduction}
One of the main differences among  blazars 
is constituted by the  position of the peak of the synchrotron
component in their spectral energy distribution (SED), namely at low
(mm--IR) or high (UV and soft X--ray) frequencies (e.g. Padovani \&
Urry 2001). 
The results of the \sax and ASCA observations  have shown that
there is a rather smooth sequence with respect to the 
peak frequencies, and in particular that this sequence is well 
extended also at high energies ($>1$ keV), in a range of physical conditions 
not previously considered.
With this respect, the \sax observations of Mkn 501 and 1ES 2344+514
have been fundamental, revealing for the first time objects with synchrotron
peak frequencies around or above 100 keV.
Such sources are of great interest also because some of them have been detected
at TeV energies by Cherenkov telescopes (Catanese \& Weekes 1999).
In these objects, X-ray and TeV observations monitor the behavior of the most
energetic electrons of the source, thus shedding light on the 
acceleration mechanism working at the most extreme conditions.
The strong correlation between TeV and X-ray emissions,
clearly evident in the 1997 flare of Mkn 501 (Pian et al. 1998, 
Aharonian et al. 1997) and in Mkn 421 (Maraschi et al. 1999, 
Krawczynski et al. 2001), together with the very rapid variability displayed
(Mkn 421 doubled its TeV flux in less than 20 min,
Gaidos et al. 1996), provides very strong 
constrains for any emission model
and a powerful tool for diagnostics.
TeV BL Lacs are also interesting because, being the only known
extragalactic sources at these energies, they allow an independent estimate 
of the extragalactic IR background (IRB), due to the absorption of high 
energy photons through $\gamma-\gamma$ collision and pair production
(Stecker et al. 1992).


With the aim to find and study more 
``extreme"\footnote{we will call ``extreme" a 
source with $\nu_{\rm peak}\gsim1$ keV} objects,  
in order to sample
more accurately the high energy branch of the peak sequence, we have 
performed an observational campaign with \sax, taking advantage
of its unique  wide energy band (0.1-100 keV).
Here we summarize the  results  on 9 sources,
observed  between 1998 and 2001, in the context of their SED.
We then consider the  impact of these objects in the determination 
of the intrinsic physical parameters governing the peak sequence,
and present a selection criterium according to which
they can be considered good targets for Cherenkov telescopes.

\section{Observational results}
The 9 observed objects have been selected from the HEAO1 survey,
the Einstein Slew Survey and the RASS catalogue.
The main selection criterium has been a very high 
$F_x/F_{\rm radio}$ ratio ($>3\times10^{-10}$ erg cm$^{-2}$
s$^{-1}$/Jy, measured at least once, in the [0.1--2.4] keV band and at 
5 GHz respectively), which corresponds roughly to $\alpha_{rx}<0.6$.
This parameter is a good indicator of the location of the synchrotron peak,
since objects with high peak frequencies are always characterized 
by low values of  $\alpha_{rx}$ (see Padovani \& Giommi 1995, Wolter et al. 1998,
Costamante et al. 2001). In addition,  a high X--ray flux ($>10^{-11}$ erg s$^{-1}$ cm$^{-2}$)
was also requested, to possibly achieve a good detection also in the 
PDS instrument.

\begin{figure}
\vspace*{-2.5cm}
\epsfysize=22cm 
\hspace*{-1.1cm} 
\epsfbox{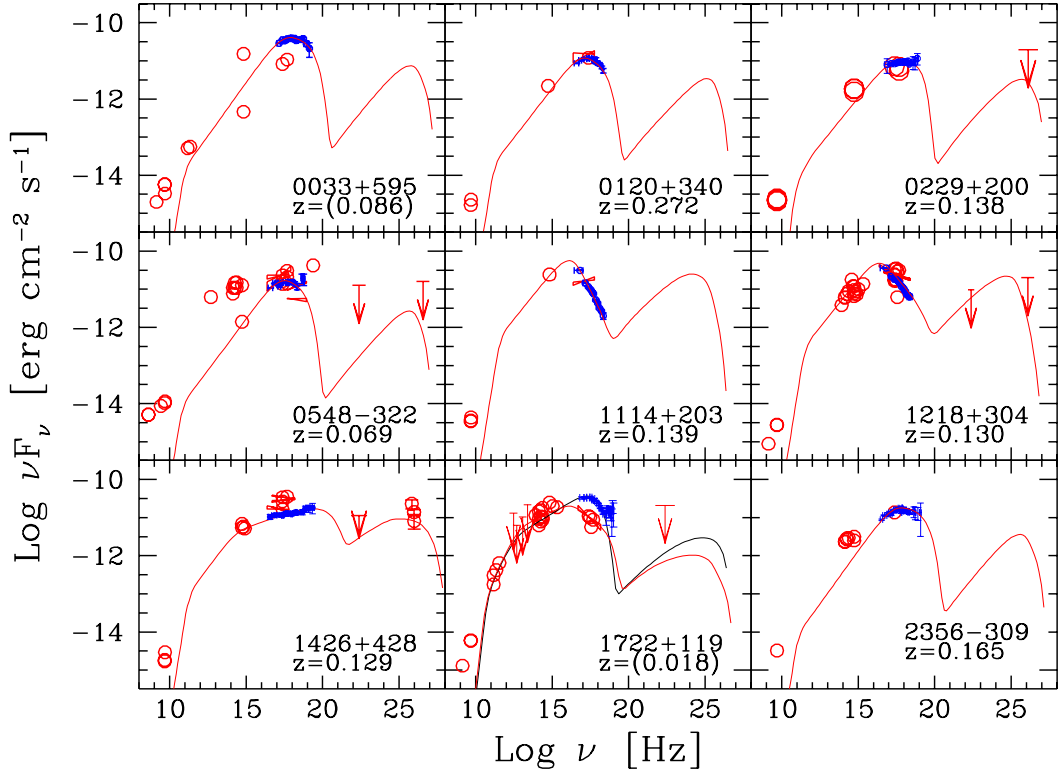}
\vspace*{-9cm}
\caption[h]{The SEDs of the 9 BL Lacs observed, made with {\it Beppo}
SAX (filled points)
and literature data (open circles, bow-ties for  
slope informations and  arrows for upper limits). Redshifts in parentheses
 are tentative (NED).  
The SEDs have been modeled with a finite injection time model
(details in  Ghisellini et al. 2002).
The TeV data for 1ES 1426+428 are from the WHIPPLE detections
(Horan et al. 2002).}
%
\end{figure}

The details of the data analysis and results are reported in
Costamante et al. 2001 (1ES 0033+595, 1ES 0120+340, PKS 0548-322, GB 1114+203,
1ES 1218+304, 1ES 1426+428, H 2356-309) and in Costamante et al. 2002
(in preparation; 1ES 0229+200 and H 1722+119).
Figure 1 shows the SEDs of all objects, assembled with the new \sax plus 
literature
data.
These observations have revealed  the ``extreme" nature of 7 
objects out of 9: for six of them (1ES 0033+595, 1ES 0120+340, 1ES 0229+200,
PKS 0548--322, H 1722+119 and H 2356--309), the synchrotron peak lies in 
the observed X-ray band, between 1 and 10 keV (as obtained from the 
broken power-law best fits), while for 1ES 1426+428 the flat X-ray spectrum 
up to 100 keV constrains the synchrotron peak to lie near or above that value. 
The other two sources, instead
(GB 1114+203 and 1ES 1218+304), displayed more "normal" HBL properties,
with a steep X-ray spectrum ($\alpha_{x}>1$) which locates the synchrotron 
peak  below the observed band.
These results confirm that the criteria adopted to select
our candidates were highly efficient (7 out of 9), and we are beginning to 
populate the high energy end of the synchrotron peak sequence.
Among these, 1ES 1426+428 is the most interesting object,
for three main reasons: 1) it is the third source ever found with 
a synchrotron peak frequency at or above 100 keV, after Mnk 501 and 1ES 2344+514;
2) it has been recently detected at TeV energies by many Cherenkov telescopes,
namely from WHIPPLE (Horan et al. 2000) and CAT (Djannati--Atai, priv. comm.)
above 300 GeV, and from
HEGRA (Aharonian et al. 2002) above 0.9 TeV. Being moreover the 
source with highest redshift detected at TeV energies ($z=0.129$), 
it is going to play a crucial role in the study of IRB;
3) it has shown a different behavior from the other ``over 100 keV" sources,
with respect to the relation between peak frequency and flux states
(see Fig. 2): the extremest state, in fact, was not associated with a 
major flaring event (at least in the observed X-ray window),
contrary to the ``higher when brighter" behavior of Mkn 501 and 1ES 2344+514.
This is revealing of a more complex relation between peak frequency 
and flux variations.

\begin{figure}
\hspace*{-0.5cm}
\epsfxsize=14cm \epsfbox{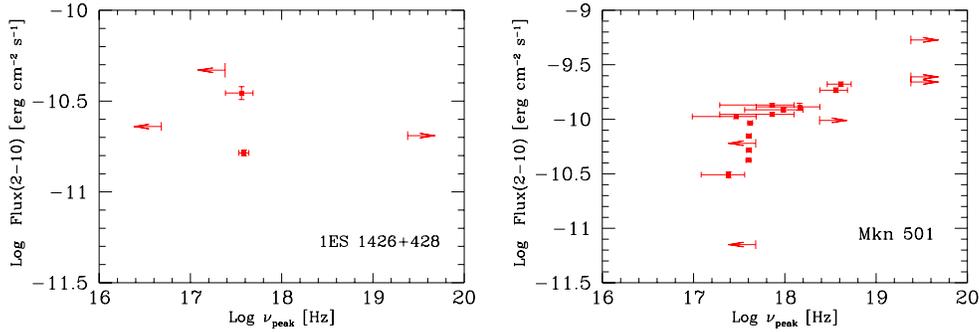} 
\caption[]{Integrated 2-10 keV flux vs. peak frequency for 
1ES 1426+428 and Mkn 501. X--ray data from literature 
(Costamante et al. 2001 and references therein; 
Pian et al. 1998, Kataoka et al. 1999, Sambruna et al. 2000, 
ASCA Tartarus database).}
\end{figure}

\section{The blazar sequence revisited}
In order to determine the intrinsic physical parameter governing 
the blazar sequence, Ghisellini et al. (1998, hereafter G98) modeled 
the SED of 
individual blazars adopting a synchrotron plus inverse Compton model,  
considering all the detected EGRET sources with $\gamma-$ray spectral shape
and redshift known. They found an inverse correlation between the energy of the 
particles emitting at the peaks of the SED, $\gamma_{\rm peak}m_{\rm e}c^2$, 
and the energy density $U$ (magnetic and radiation fields) as seen in the 
frame comoving with the emitting plasma.  
This correlation appeared to be well approximated by
$\gamma_{\rm peak} \propto U^{-0.6}$, directly implying that the radiative
cooling rate at $\gamma_{\rm peak}$ ($\propto \gamma^2_{\rm peak} U$) is
almost constant for all sources, thus suggesting a key role of the
radiative cooling process in shaping the SED.

\begin{figure}
\hskip -1cm
\epsfxsize=8cm \epsffile{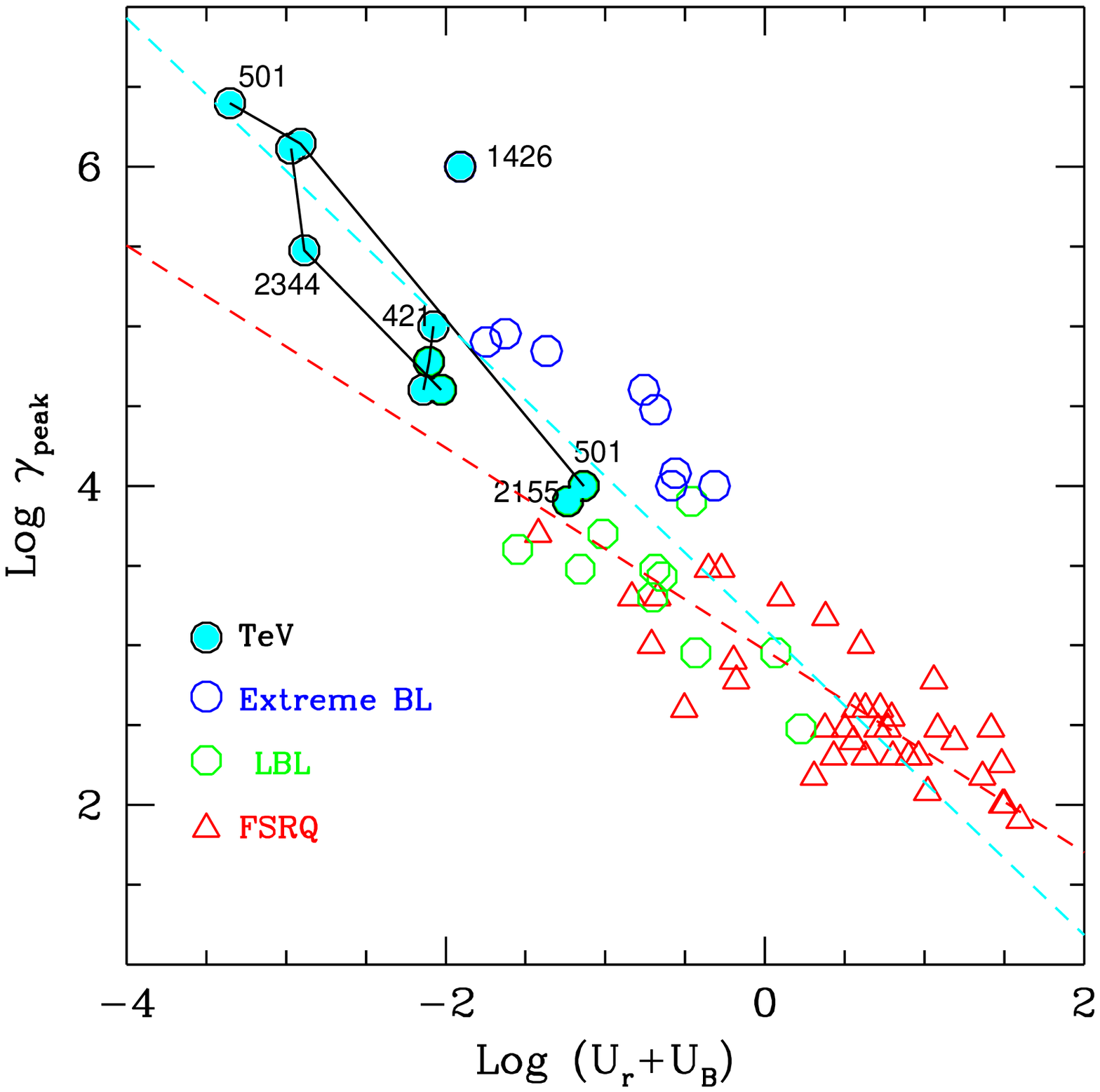} 
\vspace*{-8cm}
\hspace*{6cm}
\epsfxsize=8cm \epsffile{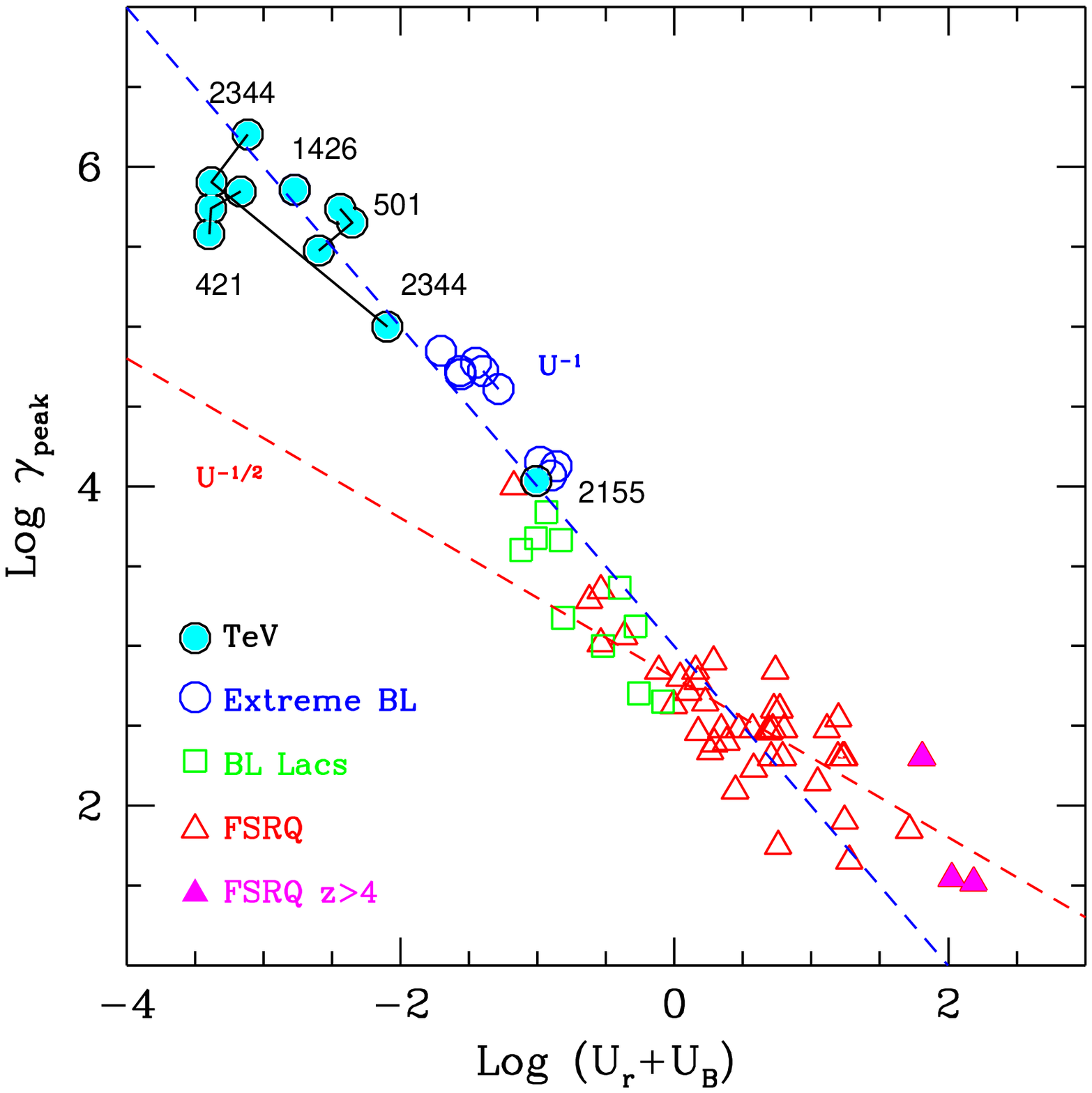}
\caption[]{The Lorentz factor of the electrons emitting at the peaks of
the SED, $\gamma_{\rm peak}$, as a function of the comoving energy density
(radiative plus magnetic). The points connected by
a line correspond to the quiescent and flaring states of the same source, 
namely Mkn 501, Mkn 421 and 1ES 2344+514, as labeled. Left: the  model used 
is the same as in G98, and the dashed lines corresponds to the linear 
correlations found in G98 and considering only the BL Lac objects,
including the extreme. Right: the model used is the finite injection time
one (see text and Ghisellini et al. 2002). The dashed lines correspond 
to $\gamma_{\rm peak} \propto U^{-1/2}$ and $\propto U^{-1}$ 
(they are not best fits).}
\end{figure}

Now that the \sax observations have provided a reasonable number
of high frequency peaked sources with sufficient data
to constrain the shape of their SED, it is  possible to explore
the  $\gamma_{\rm peak}$--$U$ correlation also at high
$\gamma_{\rm peak}$ (low $U$), in a region not  covered
by the previous sample. 

To do this, along with the sources of Sect. 2,
we (Ghisellini et al. 2002) have considered  
also other objects with good data, namely three different states for 
each of the well known sources Mkn 501, 1ES 2344+514 and Mkn 421,
and 1ES 1101-232, another extreme object recently 
discovered (Wolter et al. 2000).
The intrinsic physical parameters have been deduced modeling the SEDs
with the same model as in G98.  The results are shown in Fig. 3 (left panel),
superimposed to the original data (G98).
Fig. 3 clearly shows that $\gamma_{\rm peak}$ and $U$ 
are still significantly correlated also for the more extreme objects, 
but with a different functional dependence of 
$\gamma_{\rm peak}$ vs $U$: for
values of $\gamma_{\rm peak}$ greater than $\sim 1000$ (corresponding
to $U\lsim 1$ erg cm$^{-3}$), a new branch emerges, that can be 
approximately described by
$\gamma_{\rm peak}\propto U^{-1}$. This functional relation 
implies an approximately constant radiative cooling time
$t_{\rm cool}(\gamma_{\rm peak})$, suggesting that at low $U$
$\gamma_{\rm peak}$ might be determined by the cooling timescale approaching 
another relevant timescale of the system.
The most plausible one is the light crossing time
associated with the dimension of the emitting region, suggested also 
by the fact that quantitatively, from the parameters of the model,  
$t_{\rm cool}(\gamma_{\rm peak}) \sim R/c$.
 
A possible scenario which can naturally operate with  this timescale
is represented  by the internal shocks (see Ghisellini 1999; 
Spada et al. 2001),
in which the dissipation takes place during the collision of two shell
of fluid moving at different speeds. In this scenario, 
the particle injection time $t_{\rm inj}$ is finite, and of the order of 
the dynamical one (the crossing time of the two shells). The shape 
of the emitting particle distribution is then determined by 
the electrons having or not the time to cool within $t_{\rm inj}$:
only those  with $\gamma>\gamma_{\rm c}$ 
(defined as $t_{\rm cool}(\gamma_{\rm c})=t_{\rm inj}$) have time to cool,
and thus the particle distribution assumes a broken power-law shape
with a steep part above $\gamma_{\rm c}$ and 
the original injection slope below.
In this case, $\gamma_{\rm c}$ determines the peak of the emission
($\gamma_{\rm peak} =\gamma_{\rm c}$; details in Ghisellini et al. 2002).

The results obtained  applying this new model to the blazar SEDs
is shown in Fig. 3 (right panel), while Fig. 1 shows the SED 
of our 9 objects modeled according to it. 
The new model confirms the existence 
of two branches, obtaining in the fast cooling regime (i.e. when all the
injected particles have time to cool) the same results as in G98,
while in the slow cooling regime (low $U$) the  new branch emerges.
It's worth to note here that what we have verified for the new scenario
it's not the ``presence" of the new branch (that's a consequence of the physical 
assumptions of the new model, i.e. it's ``built-in"), but that this model
fits consistently the SEDs of all blazars, including
also the powerful ones, i.e. it provides a single and consistent scenario
with which it is possible to account for the SED of every object.

\begin{figure}
\vspace*{-1.0cm}
\epsfysize=14cm 
\hspace*{-0.7cm} 
\epsfbox{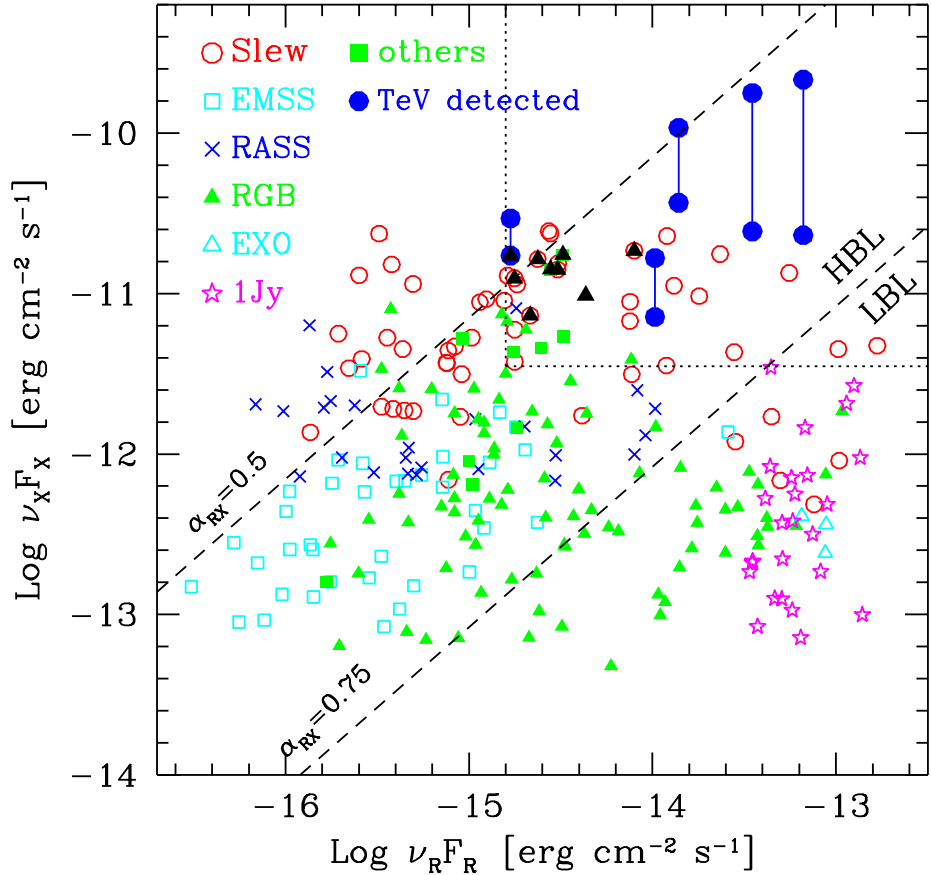}
\vspace*{-1.0cm}
\caption[h]{BL Lac objects in the radio (5 GHz) and X--ray (1 keV)
$\nu F(\nu)$ plane (details in Costamante \& Ghisellini 2001).
Sources belonging to different samples have
different symbols, as labeled.
The objects marked with filled circles are those already detected 
at TeV energies (from left to right, 1ES 1426+428, 1ES 2344+514, 
PKS 2155--304, Mkn 421, Mkn 501).
Note that for these sources we have plotted two different X--ray states,
connected by the vertical segment.
The dotted lines delimiting the rectangle are at 
$F_{\rm x}=1.46 \mu$Jy and $F_{\rm R}=31.6$ mJy, and have been drawn 
to include all the already TeV detected  objects
and sources like Mkn 501 and Mkn 421 if they were at a redshift of $\sim0.1$.
Black triangles mark the location of the 9 sources of Fig. 1.
}
\end{figure}

\section{Are extreme BL Lacs good TeV candidates ?}
Extreme BL Lacs seem at first sight the best candidates
for a copious TeV emission, since they are characterized among all blazars 
by the highest synchrotron peak frequencies, 
which is the first requirement to emit in the TeV band
(we must have many electrons energetic enough to emit at TeV energies).
Figure 4 shows the X--ray flux as a function of the radio flux 
for a large number of BL Lacs objects.
Note the locations of the already TeV--detected sources:
they are among the brightest sources in both bands. 
This is not so obvious as it may seem at first sight, since
the high peaked objects are characterized by  low values of $\alpha_{rx}$,
i.e. a large X--ray to radio flux (hence, a lower radio emission
for a given X--ray flux).
Consider also that the radio emission (at 5 GHz) must be produced
in a large region of the jet (not to be self--absorbed), much larger
than the part of the jet emitting high frequency radiation 
(as required by the very rapid variability).

We have interpreted this property in the following way 
(Costamante \& Ghisellini 2002):
to produce a large TeV flux by the Inverse Compton (IC) process
we need many electrons of random Lorentz factors $\gamma\sim 10^5$--$10^6$.
These electrons emit synchrotron photons of energies 
$h\nu = 1.5 B\gamma_5^2\delta_1$ keV.
The seed photons most effective to interact with these
electrons to produce TeV photons by the IC process
are in the IR--optical band, since photons of higher frequencies
scatter in the Klein Nishina regime.
We therefore propose that the radio flux somewhat measures the level of
the relevant seed photons.
If this is the case, then, for a given X--ray flux, sources that 
are brighter in the radio band are more likely to be TeV emitters.
And indeed the objects already detected in the TeV band are 
bright both in the radio and in the X--ray bands.

As shown in Fig. 4, in the ``rectangle" of high radio and X--ray fluxes
there are also other sources, that we therefore consider 
good candidates for TeV detection. 
Since all the 8 objects of Sect. 2, besides 1ES 1426+428, are included
in this region,  we consider them as 
good targets for Cherenkov telescopes.



\end{document}